\documentclass[onecolumn, notitlepage, PRL,%superscriptaddress
]{revtex4-1}
\usepackage{graphicx}
\usepackage{amsfonts}
\usepackage{amssymb}
\usepackage{amsthm}
\usepackage{array}
\usepackage{amsmath}
\usepackage{verbatim}
\usepackage{hyperref}
\usepackage{color}
\usepackage{bbold}
\usepackage{epstopdf}
\usepackage{mathtools}
\usepackage{enumerate}
\usepackage{subcaption}
\usepackage{color}
\usepackage{ulem}
\newtheorem*{proposition*}{Proposition}

\newtheorem{theorem}{Theorem}
\newtheorem*{theorem*}{Theorem}

\newtheorem*{corollary*}{Corollary}

\newtheorem{lemma}{Lemma}

\begin{document}
\title{Partial Boolean functions with exact quantum 1-query complexity}
\author{Guoliang Xu$^{}$}
\author{Daowen Qiu$^{}$}
\email{issqdw@mail.sysu.edu.cn (D. Qiu)}
\affiliation{ Institute of Computer Science Theory, School of Data and Computer Science, Sun Yat-sen University, Guangzhou 510006, China}
% $^2$ Instituto de Telecomunica\c{c}\~{o}es, Departamento de Matem\'{a}tica, Instituto Superior T\'{e}cnico, Av. Rovisco Pais 1049-001, Lisbon, Portugal}

\begin{abstract}
We provide two sufficient and necessary conditions to characterize any $n$-bit partial Boolean function with exact quantum 1-query complexity.
Using the first characterization, we present all $n$-bit partial Boolean functions that depend on $n$ bits and have exact quantum 1-query complexity.
Due to the second characterization, we construct a function $F$ that maps any $n$-bit partial Boolean function to some integer, and
if an $n$-bit partial Boolean function $f$ depends on $k$ bits and has exact quantum 1-query complexity, then $F(f)$  is non-positive.
In addition, we show that the number of all $n$-bit partial Boolean functions that depend on $k$ bits and have exact quantum 1-query complexity is not bigger than $n^{2}2^{2^{n-1}(1+2^{2-k})+2n^{2}}$ for all $n\geq 3$ and $k\geq 2$.
%依赖于n比特的n比特部分布尔函数，是区别于依赖于k（<n）比特的n比特部分布尔函数来说的.
%Note that the number of $n$-bit partial Boolean functions is $3^{2^{n}}$.
%Naturally, the result shows a power and advantage of the exact quantum 1-query model.
\end{abstract}

\maketitle

\section{INTRODUCTION}\label{Intro}
% Put \label in argument of \section for cross-referencing
In the field of quantum computation, the quantum query model (quantum black box model, or quantum decision tree model) is a generalization of the decision tree model \cite{Buhrman2002Complexity,BB2001Quantum,Ambainis2002adver,Childs2006Improved,H2005Lower}.
Most of famous quantum algorithms are captured by the quantum query model \cite{Nielson2000Quantum}, such as Shor's factoring algorithm \cite{Shor1994Algorithms}, Grover's unstructured search algorithm \cite{Grover1996Afast}, and many others \cite{Deutsch1985quantumtheory,Deutsch1992rapidsolution,HHL2009algorithm,quantumalgorithmzoo}.
The quantum query model can be investigated in the exact setting and the bounded-error setting \cite{Buhrman2002Complexity}.
Given an input $x\in D\subseteq\{0,1\}^{n}$ that can only be accessed through a black box by querying some bit $x_{i}$ of the input, the quantum query model computes an $n$-bit partial Boolean function $f:D\rightarrow\{0,1\}$ exactly (or with bounded-error) \cite{Buhrman2002Complexity}.
An exact quantum algorithm must always output the correct function value for all legal inputs \cite{Buhrman2002Complexity}.
If a quantum algorithm outputs the function value with a probability greater than a constant ($>\frac{1}{2}$) for all legal inputs, then the quantum algorithm is said to compute the function with bounded error.
In addition, the quantum query complexity is the decision tree complexity for the quantum machine model \cite{Buhrman2002Complexity,BB2001Quantum,Ambainis2002adver}.
Roughly speaking, the exact (or bounded-error) quantum query complexity of a Boolean function denotes the number of queries of an optimal quantum decision tree that computes the Boolean function exactly (or with bounder-error) \cite{Buhrman2002Complexity}.

In this paper, we focus on the power and advantage of the exact quantum 1-query model for partial Boolean functions.
For a partial Boolean function $f$, quantum advantages can be investigated by comparing the exact quantum query complexity $Q_{E}(f)$ and classical deterministic query complexity $D(f)$ \cite{Buhrman2002Complexity}.
Over the past decade, there have been many results on the quantum query model ~\cite{Ambainis2013Exact,ambainis2013acm,Ambainis2015qic,ambainis2016siamjc,Montanaro2015On,Ambainis2017Exact,Qiu2018Generalized,He2018Exact,Kaniewski2014Query,Montanaro2011Unbounded}.
In particular, Ambainis et al. \cite{Ambainis2015qic} proved that exact quantum algorithms have advantage for almost all Boolean functions in 2015. So far, for total Boolean functions (i.e., partial Boolean functions with $D=\{0,1\}^{n}$), Ambainis \cite{ambainis2016siamjc} in 2016 presented the best-known separation between exact quantum and classical deterministic query complexity that is a quadratic gap between exact quantum and classical deterministic query complexities, up to polylogarithmic factors.

For any partial Boolean function, the best separation between exact quantum and classical deterministic query complexity is still achieved by Deutsch-Jozsa algorithm \cite{Deutsch1992rapidsolution,Qiu2018Generalized}.
During the past five years, some main results are as follows.
In 2016, Qiu et al. \cite{Qiu2016character,Qiu2016revisit1} presented all symmetric partial Boolean functions having exact quantum 1-query complexity, and
proved that any symmetric partial Boolean function $f$ (it is a special class of partial Boolean functions) has exact quantum 1-query complexity if and only if $f$ can be computed by the Deutsch-Jozsa algorithm \cite{Deutsch1992rapidsolution}.
In the same year, Aaronson et al. \cite{Aaronson2016ccc} showed an equivalence
%(Arunachalam et al. \cite{Arunachalam2019siam} gave a simple proof of the equivalence in 2019)
between quantum 1-query algorithms and bounded quadratic polynomials in the bounded-error setting.
In 2019, Arunachalam et al. \cite{Arunachalam2019siam} proved a characterization of quantum $t$-query algorithms in terms of the unit ball of a space of degree-$(2t)$ polynomials.

Recently, Chen et al. \cite{Chen2020character} proved that a total Boolean function can be computed exactly by a quantum 1-query algorithm if and only if $f(x)=x_{i1}$ or $x_{i1} \oplus x_{i2}$ (up to  equivalence). 
Indeed, the proof of necessity in \cite{Chen2020character} is the same method as the proof of Theorem 11 in  \cite{Qiu2016character,Qiu2016revisit1}.
Very soon  Mukherjee et al. \cite{Mukherjee2020Classical} noticed that the main result of \cite{Chen2020character} is the same as  a result by Montanaro, Jozsa, and Mitchison \cite{Montanaro2015On}.

However, note that all these results are not applicable to the exact quantum 1-query model for all partial Boolean functions, so we investigate the following two problems.
\begin{itemize}
\item [(1)] The partial Boolean function can be regarded as a generalization of the total Boolean function. Actually, Deutsch's algorithm \cite{Deutsch1985quantumtheory} computes a two-bit partial (also total) Boolean function using one query. And, both the extension of Deutsch's problem (computed by Deutsch-Jozsa algorithm \cite{Deutsch1992rapidsolution}) and a generalized Deutsch-Jozsa problem in Ref.~\cite{Qiu2018Generalized} are described by even $n$-bit partial (not total) Boolean functions. Naturally, {\bf what is the characterization of partial Boolean functions with exact quantum 1-query complexity?}
\item [(2)] In the field of quantum computation, it is a fundamental and interesting subject to evaluate the computational power of the quantum 1-query model, and is also critical for discovering quantum advantage. Specifically, the number of partial Boolean functions with exact quantum 1-query complexity shows the power and advantage of the exact quantum 1-query model. So, {\bf how many partial Boolean functions can be computed exactly by quantum 1-query algorithms?}
\end{itemize}

The rest of the paper is organized as follows. In Section \ref{preli}, we introduce some basis notations and the related knowledge. Then, we give and prove three main results in Section \ref{mainresults}.
Finally, the conclusion is presented in Section \ref{Concl}. For the sake of brevity and readability, all proofs of lemmas in this paper are showed in Appendixes.

\section{PRELIMINARIES}\label{preli}

In this section, we introduce some basic notations and recall some basic knowledge of partial Boolean functions and the exact quantum query model. For the details, we can refer to Refs. \cite{Simon1983A,Nielson2000Quantum,Buhrman2002Complexity,WolfNondeterministic2003,Montanaro2011Unbounded,Qiu2016character}.

As usual, notations $N$, $R$, and $C$ denote the sets of integer numbers,  real numbers,  and  complex numbers, respectively.
In particular, we will always use the notation $D$ (or promised set) to denote a subset of $\{0,1\}^{n}$.
For any input $x=x_{1}x_{2}\cdots x_{n}\in D$, the Hamming weight (number of 1s) of $x$ is denoted by $|x|$.
Given a real number set $S$, the notation $\max S$  denotes the maximum in $S$ and the notation $\min S$ denotes the minimum in $S$. For any finite set $S$, the notation $|S|$ denotes the number of elements in $S$.
For a complex matrix $A$, $A^{T}$ is the transpose of the matrix $A$, and $A^{\dag}=(A^{T})^{*}$ is the conjugate transpose of the matrix $A$. Obviously, $A^{\dag}=A^{T}$ for any real matrix $A$. Furthermore, the notation $|a\rangle$ is usually used to denote a column vector which is labeled by the notation $a$ and $\langle a|=(|a\rangle)^{\dag}$ is a row vector.

In this paper, we mainly concern partial functions $f: D\rightarrow C$, $f: D\rightarrow R$ and $f: D\rightarrow \{0,1\}$.
In general, these functions can be given by a $2^{n}$-dimensional vector $(f(0),f(1),$ $\cdots,f(x),$ $\cdots,f(2^{n}-1))^{T}$ whose entry $f(x)$ is the notation $*$ for any undefined input $x\in\{0,1\}^{n}/D$.
For example, the Boolean function $f$ computed by Deutsch's algorithm \cite{Deutsch1985quantumtheory} can be given by $(f(00),f(01),f(10),f(11))=(1,0,0,1)$.
Sometimes, we also use a two-tuple $(\{x|f(x)=0\},\{x|f(x)=1\})$ to give a certain partial Boolean function $f: D\rightarrow\{0,1\}$.
For example, the even $n$-bit partial Boolean function $f$ computed by Deutsch-Jozsa algorithm \cite{Deutsch1992rapidsolution} can be given by $(\{x|f(x)=0\},\{x|f(x)=1\})=$ $(\{x|x\in\{0,1\}^{n},|x|=0,n\},\{x|x\in\{0,1\}^{n},|x|=\frac{n}{2}\})$.
In addition, the notation $\bar{f}$ denotes the negation of $f$ and $\bar{f}(x)=1\oplus f(x)$ for all $x\in D$.

In order to represent these functions, we need to use two monomials $X_{S}=\prod_{i\in S}x_{i}$ and $(-1)^{S\cdot x}=\prod_{i\in S}(-1)^{x_{i}}$ \cite{Buhrman2002Complexity,Montanaro2011Unbounded,WolfNondeterministic2003}.
In particular, $X_{\emptyset}=(-1)^{\emptyset\cdot x}=1$.
And, the set $\{X_{S}|S\subseteq\{1,2,\cdots,n\}\}$ is usually called as the polynomial basis and the set $\{(-1)^{S\cdot x}|S\subseteq\{1,2,\cdots,n\}\}$ is usually called as Fourier basis \cite{Buhrman2002Complexity,WolfNondeterministic2003,Montanaro2011Unbounded}.
If a function $p:R^{n}\rightarrow C$ can be written as $\Sigma_{S}\alpha_{S}X_{S}$ for some complex numbers $\alpha_{S}$, then the function $p$ is called a multilinear polynomial \cite{Buhrman2002Complexity}.
Meanwhile, the degree of the multilinear polynomial $p$ is defined by $\deg(p)=\max\{|S||\alpha_{S}\neq 0\}$.
For any partial function $f:D\rightarrow C$, a multilinear polynomial $p(x)$ represents $f$ if and only if $p(x)=f(x)$ for all $x\in D$ \cite{Buhrman2002Complexity,NisanOn}.
Unlike total functions $f:\{0,1\}^{n}\rightarrow C$, the multilinear representation of a partial (not total) function $f:D\rightarrow C$ is usually not unique.
Thus, the degree of a partial (or total) function $f:D\rightarrow C$ can be defined by $\deg(f)=\min\{\deg(p)|p~\textrm{represents}~f\}$.

In the quantum query model, for every input $x\in D$, the quantum black box $O_{x}$ can be described as a unitary operator which is defined by
\begin{equation}\label{actionofox}
O_{x}|i,j'\rangle=\left\{
\begin{array}{lcl}
(-1)^{x_{i}}|i,j'\rangle, & & \textrm{if}~~i\in\{1,2,\cdots,n\},\\
|0,j'\rangle, & & \textrm{if}~~i=0.
\end{array} \right.
\end{equation}
Since the index $j\in\{0,1,\cdots,n'-1\}$ can be mapped onto $(i,j')$ in one-to-one way, $j$ can be labeled as $(i,j')$.
Here, the integer number $n'$ denotes the number of basis states in the quantum query algorithm, the integer number $i\in\{0,1,2,\cdots,n\}$ is the query-part and the label $j'$ is the other-part.
By assuming no ambiguity exists, the bracket can be omitted in Dirac notation and $|j\rangle=|(i,j')\rangle$ can be written as $|i,j'\rangle$.
Then, a quantum $t$-query algorithm can be determined by an initial state $|\psi_{0}\rangle$ and a sequence of unitary transformations $U_{0},O_{x},U_{1},O_{x},\cdots,O_{x},U_{t}$ followed by a measurement, where $t+1$ unitary operators $U_{0},U_{1},\cdots,U_{t}$ are independent of the input \cite{Nielson2000Quantum,Buhrman2002Complexity}.

\section{Main results}\label{mainresults}

In order to show our results, we need some notations in the following definition.
\newtheorem{myDef}{Definition}
\begin{myDef}\label{def4}
The $2^{n}\times 2^{n}$ matrix
\begin{equation}\label{}
G_{n}=(|P(0)\rangle,\cdots,|P(x)\rangle,\cdots,|P(2^{n}-1)\rangle).
\end{equation}
Here, the polynomial basis vector
\begin{equation}\label{}
|P(x)\rangle{=}(X_{\emptyset},\cdots, X_{S}, \cdots, X_{\{1,\cdots,n\}})^{T}
\end{equation}
is a $2^{n}$-dimensional vector function on the variable $x=x_{1}x_{2}\cdots x_{n}\in \{0,1\}^{n}$ where every number set $S$ in $|P(x)\rangle$ is a subset of $\{1,2,\cdots,n\}$.

The order of $S$ in $|P(x)\rangle$ and the order of $|P(x)\rangle$ in $G_{n}$ are arranged as follows.
For every $m\in\{0,1,2,\cdots,n\}$, we first sort all elements of the set $\{X_{S}||S|=m\}$ into a basic block vector function $(\cdots,X_{S},\cdots)^{T}_{|S|=m}$ based on a pre-fixed order of $S$ (the order remains unchanged in all related discussions). Then, based on the ascending order of $m\in\{0,1,2,\cdots,n\}$, we sort these basic blocks into $|P(x)\rangle$ from top to bottom. Finally, using the established order of $S$, $|P(x)\rangle$ in $G_{n}$ are sorted from left to right based on the mapping $S\rightarrow x=x_{1}x_{2}\cdots x_{n}$ where $x_{i}=1$ if and only if $i\in S$.

Given a positive integer $m\in\{1,2,\cdots,n\}$, the $\sum^{m}_{r=0}\binom{n}{r}$-dimensional vector function
\begin{equation}\label{}
|P(x)\rangle_{m}{=}(X_{\emptyset}, \cdots, X_{S}, \cdots, X_{\{1,\cdots,m\}})^{T}
\end{equation}
denotes a sub-vector of $|P(x)\rangle$ where every number set $S\subseteq\{1,2,\cdots,n\}$ satisfying $|S|\leq m$. Similarly, the Fourier basis vector $|F(x)\rangle$ and $|F(x)\rangle_{m}$ can be defined by replacing every monomial $X_{S}$ in $|P(x)\rangle$ and $|P(x)\rangle_{m}$ with the monomial $(-1)^{S\cdot x}$, respectively.
For an $n$-bit partial Boolean function $f: D\rightarrow\{0,1\}$, let $G_{f}(m)$ be the sub-matrix $(\cdots,|P(x)\rangle_{m},\cdots)_{x\in D}$ (the matrix consists of all column vectors $|P(x)\rangle_{m}$ satisfying $x\in D$) which is extracted from $G_{n}$, and let $G_{f}(m,b)$ $(m\leq n)$ be the sub-matrix $(\cdots,|P(x)\rangle_{m},\cdots)_{f(x)=b}$ (the matrix consists of all column vectors $|P(x)\rangle_{m}$ satisfying $f(x)=b$) which can also be extracted from $G_{n}$.
\hfill $\Box$
\end{myDef}
Using Definition \ref{def4}, any $n$-bit partial function $f:D\rightarrow C$ can be represented by $(|\alpha_{f}\rangle)^{\dag}|P(x)\rangle$ where $(|\alpha_{f}\rangle)^{\dag}$ $=$ $(\cdots,\alpha_{S},\cdots)$ is a polynomial coefficient vector.
When the degree of $f$ is $d$, $f$ can also be represented by $(|\alpha_{f}\rangle_{d})^{\dag}|P(x)\rangle_{d}$. Here, it should be emphasized that the polynomial coefficient vector can be transformed to the Fourier coefficient vector using an invertible matrix $A$. Also, $A$ can be determined by equations $x_{i}=-\frac{1}{2}(-1)^{x_{i}}+\frac{1}{2}$ for all $i\in \{1,2,\cdots,n\}$.
Inspired by proofs of Theorem 8 in Ref.~\cite{Qiu2016character} and Lemma 1 in Ref.~\cite{Chen2020character}, the first characterization is presented in the following.
\begin{theorem}\label{the0}
An $n$-bit non-constant partial Boolean function $f:$ $D\rightarrow$ $\{0,1\}$ can be computed exactly by a quantum 1-query algorithm, if and only if there exists a non-negative solution $\vec{\beta}=$($\beta_{0},$$\beta_{1},$$\beta_{2},$$\cdots,$$\beta_{n})^{T}$ of equations $\beta_{0}+$$\beta_{1}+$$\beta_{2}+$$\cdots+$$\beta_{n}$=1 and $(|F(x\oplus y)\rangle_{1})^{T}\vec{\beta}=0$ for all $x\in\{x|f(x)=0\}$ and $y\in\{x|f(x)=1\}$. \hfill $\Box$
\end{theorem}
Then, using the following definition \cite{Simon1983A,Ambainis2015qic,Chen2020character} (the background of this definition can be seen in Appendix), we will get a result as a by-product of Theorem \ref{the0}.
\begin{myDef}\label{def441}
\cite{Simon1983A,Ambainis2015qic,Chen2020character}. An $n$-bit partial Boolean function $f: D\rightarrow\{0,1\}$ is said to depend on $k$ ($\leq n$) bits, if
$k$ is the minimum number of variables in all multilinear polynomials representing $f$.
%$|\bigcup_{S}\{S|\alpha_{S}\neq0\}|\geq k$ for any multilinear polynomial representation $\Sigma_{S}\alpha_{S}X_{S}$ of $f$ and the lower bound $k$ can be reached.
%表示f的所有多线性多项式中变量最少的那个多项式的变量个数。按原始定义，（f依赖于第i比特，意味着存在一对输入仅第i比特不同但函数值不同）f依赖于n比特可以自然的定义为依赖于所有比特。然而，对于部分布尔函数，这有一点不同。因为我们可以确定一些布尔函数，其不依赖于任意一比特，如D-J问题(2k比特布尔函数)，其任意两个输入都不可能仅仅只有一比特不同，此函数不是常量函数，其函数值可以由任意k+1比特确定（只要此k+1比特中既有0又有1那么即可确定此输入为平衡的，否则为常量的）。所以，D-J问题中的函数不依赖于任意一比特，但依赖于k+1比特。
\hfill $\Box$
\end{myDef}
Actually, the two-bit total Boolean function computed by Deutsch's algorithm \cite{Deutsch1985quantumtheory} depends on two bits, and the even $n$-bit partial Boolean function computed by Deutsch-Jozsa algorithm \cite{Deutsch1992rapidsolution} depends on $\frac{n}{2}+1$ bits.
Now, the result is stated as follows.
\begin{theorem}\label{the0.1}
For any $n$-bit partial Boolean function $f: D\rightarrow\{0,1\}$ depending on $n$ bits, $f$ can be computed exactly by a quantum 1-query algorithm, if and only if $f(x)=x_{1}$ or $1\oplus x_{1}$ with $D=\{0,1\}$, or $f(x)=x_{1}\oplus x_{2}$ or $1\oplus x_{1}\oplus x_{2}$ with $D\in\{E|E\subseteq\{0,1\}^{2},|E|\in\{3,4\}\}$. \hfill $\Box$
\end{theorem}
Theorem \ref{the0.1} tells us that there exists no an unknown $n$-bit partial Boolean function which depends on $n$ bits and has exact quantum 1-query complexity.
Here, the number of all $n$-bit partial Boolean functions depending on $n$ bits is not less than $2\times3^{2^{n}-n-1}$ (this fact can be seen in Lemma \ref{lem8}).
In contrast, the number of all $n$-bit total Boolean functions  is $2^{2^{n}}$. The number of all $n$-bit symmetric partial Boolean functions (investigated by  Qiu et al. \cite{Qiu2016character,Qiu2016revisit1}) is $3^{n}$.

In order to give the second characterization, the following notation is necessary.
\begin{myDef}\label{def5}
\cite{PowersAn,Lasserre2006A,Lasserre2006C,Papachristodoulou2013SOSTOOLS,Blekherman2016Sums,Lee2016On}. For an $n$-bit partial Boolean function $f: D\rightarrow\{0,1\}$ and a $(p+q)\times(\sum^{m}_{i=0}\binom{n}{i})$ complex matrix
\begin{equation}\label{representsos}
[\alpha_{f}]{=}\begin{bmatrix}
\begin{smallmatrix}
\langle\alpha^{1}|\\
\vdots \\
\langle\alpha^{p}|\\
\langle\alpha^{p+1}|\\
\vdots \\
\langle\alpha^{p+q}|\\
\end{smallmatrix}
\end{bmatrix}{=}[\cdots,|\alpha_{S}\rangle,\cdots],
\end{equation}
where $S\subseteq\{1,2,\cdots,n\}$ satisfying $|S|\leq m$, if
\begin{equation}\label{final11}
\left \{
 \begin{aligned}
&f(x)=\sum^{p}_{l=1}|\langle\alpha^{l}|F(x)\rangle_{m}|^{2}, &x\in D,\\
&\bar{f}(x){=}1-\sum^{p}_{l=1}|\langle\alpha^{l}|F(x)\rangle_{m}|^{2}, &x\in\{0,1\}^{n},\\
&\bar{f}(x){=}\sum^{p+q}_{l=p+1}|\langle\alpha^{l}|F(x)\rangle_{m}|^{2}, &x\in D,
 \end{aligned}
\right.
\end{equation}
then the matrix $[\alpha_{f}]$ is called a degree-$m$ SOS complex (Fourier polynomials) representation matrix of $f$ and $\bar{f}$.
%实数多项式的实平方和表示主要在优化理论中有研究，由于有些实数多项式有可能没有平方和表示。即使有平方和表示的也基本是唯一的，所以还是可以找到的，或者确定不能找到。
%布尔输入的实函数，平方和表示可以是实多线性多项式表示，这在最近的文献中都可以看到。也有平方和度的定义。
%由以上，本文联想到部分布尔函数都有平方和实多线性多项式表示并且不唯一。由于查询算法中量子态可以用复多线性多项式表示，因而这里定义时用复多线性多项式。
Here, the degree of an SOS complex representation is the maximum degree of these complex Fourier polynomials in representation.
\hfill $\Box$
\end{myDef}
Clearly, if there exists a pair of degree-1 SOS complex representation of $f$ and $\bar{f}$, then an SOS complex representation matrix of $f$ and $\bar{f}$ is in the form of $[|\alpha_{\emptyset}\rangle,$ $|\alpha_{\{1\}}\rangle,$ $|\alpha_{\{2\}}\rangle,$ $\cdots,$ $|\alpha_{\{n\}}\rangle]$.
By Definition \ref{def5}, the second characterization is presented as follows.
\begin{theorem}\label{the2}
For any $n$-bit non-constant partial Boolean function $f:$ $D\rightarrow$ $\{0,1\}$, $f$ can be computed exactly by a quantum 1-query algorithm, if and only if there exists a degree-1 SOS complex representation matrix $[\alpha_{f}]$ of $f$ and $\bar{f}$ such that
\begin{equation}\label{reprematrix3}
[\alpha_{f}]^{\dag}[\alpha_{f}]=diag(u_{0},u_{1},u_{2},\cdots,u_{n}).
\end{equation}
\end{theorem}
$\mathbf{Remark~1.}$
For a partial Boolean function $f$, we can get a pair of SOS real representation of $f$ and $\bar{f}$ first, and then transform it into a proper SOS complex representation matrix.
Since it is possible to get a pair of SOS real representation for very small (partial) Boolean functions \cite{PowersAn,Lasserre2006A,Lasserre2006C,Papachristodoulou2013SOSTOOLS,Blekherman2016Sums,Lee2016On}, Theorem \ref{the2} can be tested on very small partial Boolean functions.
%对布尔函数来说，一对合适的（可以满足定理2的条件的）平方和实表示很难得到（据我们的计算，这源于二次方程很难得到精确解）。但是输入很小的布尔函数我们的确可以通过传统的方法具体问题具体分析求解，还是可以得到的。对于较大输入的，我们并不知道有什么好的方法可以解，除非解特别明显可以观察到。
Remark 1 is finished. \hfill $\Box$

Now, we give the fourth result in the following.
\begin{theorem}\label{the1}
For any $n$-bit non-constant partial Boolean function $f: D\rightarrow\{0,1\}$, if $f$ depends on $k$ bits and can be computed exactly by a quantum 1-query algorithm, then
\begin{equation}\label{the1equation2}
rank(G_{f}(1,0)),rank(G_{f}(1,1))\in\{1,2,\cdots,n\}
\end{equation}
and
\begin{equation}\label{the1equation1}
rank(G_{f}(1,0))+rank(G_{f}(1,1))-(2n+2-k)\leq 0.
\end{equation}
%In particular, if $f$ depends on $n$ bits, then
%$rank(G_{f}(1,0))$ $+rank(G_{f}(1,1))$ $\in\{n+1,n+2\}$.
\end{theorem}
$\mathbf{Remark~2.}$ The inverse direction of Theorem \ref{the1} is not always hold. For example, a three-bit partial Boolean function $f$ given by $(\{x|f(x)=0\},\{x|f(x)=1\})=$ $(\{x|x\in\{0,1\}^{3},|x|=0\},\{x|x\in\{0,1\}^{3},|x|=1\})$.
Here, we can know that $rank(G_{f}(1,0))=1$ and $rank(G_{f}(1,1))=3$. However, using Theorem 10 in Ref.~\cite{Qiu2016character}, $Q_{E}(f)\geq 2$.
Remark 2 is finished.\hfill $\Box$

Finally, the fifth result is in the following.
\begin{theorem}\label{the3}
Let $N_{1}(n,k)$ be the number of all $n$-bit partial Boolean functions which depend on $k$ bits and have exact quantum 1-query complexity. If $n\geq 3$ and $k\geq 2$, then $N_{1}(n,k)\leq n^{2}2^{2^{n-1}(1+2^{2-k})+2n^{2}}$.
\end{theorem}

$\mathbf{Remark~3.}$
In contrast, the number $3^{2^{n}}$ is the number of all $n$-bit partial Boolean functions  that each $n$-bit partial Boolean function corresponds to a string $f(0)f(1)\cdots f(2^{n}-1)\in\{0,1,*\}^{2^{n}}$.
In fact, the exact quantum query complexity of any $n$-bit partial Boolean function is in the set $\{0,1,2,\cdots,n\}$, which implies that $\max\{N_{j}(n,k)|j,k\in\{0,1,2,\cdots,n\}\}\geq\frac{3^{2^{n}}}{(n+1)^{2}}$.
Here, the notation $N_{j}(n,k)$ denotes the number of $n$-bit partial Boolean functions which depend on $k$ bits and have exact quantum $j$-query complexity.
Thus, all $n$-bit partial Boolean functions with exact quantum 1-query complexity only make up a very tiny proportion of all $n$-bit Boolean functions.
Remark 3 is finished. \hfill $\Box$

Finally, corresponding to Fact 1 in Ref.~\cite{Qiu2016character}, the following Fact 2 is also applicable to all partial Boolean functions, as a common quantum 1-query algorithm computes the two partial Boolean functions.

$\mathbf{Fact~2.}$ For any two partial Boolean functions $f$ and $g$ satisfying $\{x|g(x)=0\}\subseteq\{x|f(x)=0\}$ and $\{x|g(x)=1\}\subseteq\{x|f(x)=1\}$, if $f$ can be computed exactly by a quantum 1-query algorithm, then $g$ can also be computed exactly by this quantum 1-query algorithm. \hfill $\Box$

\subsection{Proof of Theorem \ref{the0}}\label{proof0}

\begin{proof}
$\Rightarrow)$. Since the algorithm is exact, the quantum state $U_{1}O_{x}U_{0}|\psi\rangle$ for all $x\in\{x|f(x)=0\}$ must be orthogonal to the quantum state $U_{1}O_{y}U_{0}|\psi\rangle$ for all $y\in\{x|f(x)=1\}$. Since the unitary operator $U_{1}$ preserves the inner product of any two complex vectors, the quantum state $O_{x}U_{0}|\psi\rangle$ for all $x\in\{x|f(x)=0\}$ must be orthogonal to the quantum state $O_{y}U_{0}|\psi\rangle$ for all $y\in\{x|f(x)=1\}$. For any state $U_{0}|\psi\rangle=\sum_{i,j'}\alpha_{i,j'}|i,j'\rangle,$ note that
\begin{equation}\label{oxu0inti}
O_{x}U_{0}|\psi\rangle=\sum_{j'}\alpha_{0,j'}|0,j'\rangle+\sum_{i,j'}\alpha_{i,j'}(-1)^{x_{i}}|i,j'\rangle
\end{equation}
for all $x\in D$. Then, for all $x\in\{x|f(x)=0\}~\textrm{and}~y\in\{x|f(x)=1\}$, the inner product
$(O_{x}U_{0}|\psi\rangle)^{\dag}$ $O_{y}U_{0}|\psi\rangle$ $=(\beta_{0},\beta_{1},$ $\cdots,\beta_{n})|F(x\oplus y)\rangle_{1}$ $=0$ where $\beta_{i}=\sum_{j'}|\alpha_{i,j'}|^{2}$ for all $i\in\{0,1,2,\cdots,n\}$. In other word, there exists at least one non-negative solution $\vec{\beta}=$($\beta_{0},$$\beta_{1},$$\beta_{2},$$\cdots,$$\beta_{n})^{T}$ of equations $\beta_{0}+$$\beta_{1}+$$\beta_{2}+$$\cdots+$$\beta_{n}$=1 such that $(|F(x\oplus y)\rangle_{1})^{T}\vec{\beta}=0$ for all $x\in\{x|f(x)=0\}$ and $y\in\{x|f(x)=1\}$.

$\Leftarrow)$. For a non-negative solution $\vec{\beta}=$($\beta_{0},$ $\beta_{1},$ $\beta_{2},$ $\cdots,$ $\beta_{n})^{T}$ of equations $\beta_{0}+$$\beta_{1}+$$\beta_{2}+$$\cdots+$$\beta_{n}$=1 and $(|F(x\oplus y)\rangle_{1})^{T}\vec{\beta}=0$ for all $x\in\{x|f(x)=0\}$ and $y\in\{x|f(x)=0\}$, if we set $\sum_{j'}|\alpha_{i,j'}|^{2}=\beta_{i}$ for all $i\in\{0,1,2,\cdots,n\}$ in the state $U_{0}|\psi\rangle=\sum_{i,j'}\alpha_{i,j'}|i,j'\rangle,$ then the inner product
\begin{equation}\label{}
 \begin{split}
(U_{1}O_{x}U_{0}|\psi\rangle)^{\dag}U_{1}O_{y}U_{0}|\psi\rangle=0
 \end{split}
\end{equation}
for all $x\in\{x|f(x)=0\}~\textrm{and}~y\in\{x|f(x)=1\}$.
By Gram-Schmidt orthogonalization, we can get an orthonormal base of vectors set $\{U_{1}O_{x}U_{0}|\psi\rangle|f(x)=0\}$ and an orthonormal base of vectors set $\{U_{1}O_{x}U_{0}|\psi\rangle|f(x)=1\}$, respectively. By using the measurement consisting of the two orthonormal base set, the quantum 1-query algorithm computes $f$ exactly. Thus, Theorem \ref{the0} has been proved.
\end{proof}

\subsection{Proof of Theorem \ref{the0.1}}\label{proof0.1}
First, it should be pointed out that the number of all $n$-bit partial Boolean functions depending on $n$ bits is quite big. This fact is implied by the following lemma.
\begin{lemma}\label{lem8}
Let $N(n)$ $(n\geq1)$ denote the number of all $n$-bit partial Boolean functions depending on $n$ bits. Then,
$N(n)\geq2\times3^{2^{n}-n-1}.$
\end{lemma}
Next, Theorem \ref{the0.1} is proved in the following.
\begin{proof}
$\Rightarrow)$. For any $n$-bit partial Boolean function $f: D\rightarrow\{0,1\}$ and $k\in\{1,2,\cdots,n\}$, any multilinear polynomial representation of $f$ can be written as $f(x)=$ $x_{k}q_{1}(x_{1},x_{2},\cdots,$$x_{n})+$$q_{2}(x_{1},x_{2},\cdots,$$x_{n})$ where $q_{1}(x_{1},x_{2},\cdots,$$x_{n})$ and $q_{2}(x_{1},x_{2},\cdots,$$x_{n})$ are two multilinear polynomials on variables $x_{1},\cdots,x_{k-1},x_{k+1},\cdots,x_{n}$.
With a trivial argument, if $f$ depends on $n$ bits, then there must exist at least $n$ input pairs $(X_{1},X^{\{1\}}_{1})$, $(X_{2},X^{\{2\}}_{2})$, $\cdots,$ $(X_{n},X^{\{n\}}_{n})$ such that $1\oplus f(X_{k})=f(X^{\{k\}}_{k})\in\{0,1\}$ for all $k\in\{1,2,\cdots,n\}$.
Here, $X^{\{k\}}_{k}$ is the same as $X_{k}$ except for the $k$-th bit being flipped.
By Theorem \ref{the0}, if $f$ can be computed exactly by a quantum 1-query algorithm, then there exists a non-negative solution $\vec{\beta}=$($\beta_{0},$$\beta_{1},$$\beta_{2},$$\cdots,$$\beta_{n})^{T}$ of equations $\beta_{0}+$$\beta_{1}+$$\beta_{2}+$$\cdots+$$\beta_{n}$=1 such that $(|F(X_{k}\oplus X^{\{k\}}_{k})\rangle_{1})^{T}\vec{\beta}=0$ for all $k\in\{1,2,\cdots,n\}$. Thus, $2\beta_{k}=1$ for all $k\in\{1,2,\cdots,n\}$ which implies that $n=2$ with $\vec{\beta}=(0,\frac{1}{2},\frac{1}{2})$ or $n=1$ with $\vec{\beta}=(\frac{1}{2},\frac{1}{2})$.

The case $n=1$ is trivial, and $f$ can be given by $(f(0),f(1))=$ $(0,1)$ or $(1,0)$.
For the case $n=2$, the unique non-negative solution $\vec{\beta}=(0,\frac{1}{2},\frac{1}{2})$ implies that $\frac{1}{2}(-1)^{x_{1}\oplus y_{1}}+\frac{1}{2}(-1)^{x_{2}\oplus y_{2}}=0$ for all $x\in\{x|f(x)=0\}$ and $y\in\{x|f(x)=1\}$.
Then, $x_{1}\oplus x_{2}\neq y_{1}\oplus y_{2}$ for all $x\in\{x|f(x)=0\}$ and $y\in\{x|f(x)=1\}$.
This result implies that $f(x)=x_{1}\oplus x_{2}$ or $1\oplus x_{1}\oplus x_{2}$.
Meanwhile, since $f: D\rightarrow\{0,1\}$ is a two-bit partial Boolean function depending on two bits, $|D|\in\{3,4\}$.

$\Leftarrow)$. This direction is trivial. Thus, Theorem \ref{the0.1} has been proved.
\end{proof}

\subsection{Proof of Theorem \ref{the2}}\label{proof1}

First, the following lemma follows the discussion of Lemma 7 and Theorem 17 in \cite{Buhrman2002Complexity}.
\begin{lemma}\label{lem1}
~\cite{Buhrman2002Complexity}.
If there exists an exact quantum 1-query algorithm computing an $n$-bit partial Boolean function $f: D\rightarrow\{0,1\}$, then there must exist a pair of degree-1 SOS complex (multilinear polynomials) representation of $f$ and $\bar{f}$.
\end{lemma}

 If we knows the final state of a quantum query algorithm computing $f$ well, then with the proof of Lemma \ref{lem1}, the measurement can be determined naturally.
%实际上，我们对末态的了解就是需要知道哪些基态的振幅构成f的平方和表示, 另一些基态的振幅构成f的否定的平方和表示。如果我们知道了这些，那么测量到对应到f的平方和表示的那些基态, 就输出1. 否则，就输出0. 这一点可以通过引理1的证明看出来。

Then, a trivial matrix representation of a state in a quantum query algorithm is introduced as follows. Similar to represent a quantum state with a unit vector, every state in a quantum query algorithm will be identified with the representation matrix.
\begin{myDef}\label{def6}
For a state $U_{m}O_{x}U_{m-1}$ $\cdots $ $U_{0}|\psi_{0}\rangle$ in a quantum query algorithm and an $n'\times\left(\sum^{m'}_{i=0}\binom{n}{i}\right)$ ($m'\leq m$) matrix
\newcommand{\udots}{\mathinner{\mskip1mu\raise1pt\vbox{\kern7pt\hbox{.}}
        \mskip2mu\raise4pt\hbox{.}\mskip2mu\raise7pt\hbox{.}\mskip1mu}}
\begin{equation}\label{matrixrepresentation}
\left[\alpha^{S}_{j,m}\right]=\begin{bmatrix}
\begin{smallmatrix}
 \ddots &  & \udots\\
         &\alpha^{S}_{j,m}& \\
 \udots &   & \ddots
\end{smallmatrix}
\end{bmatrix}=[\cdots,|\alpha_{S,m}\rangle,\cdots]
\end{equation}
where $S\subseteq\{1,2,\cdots,n\}$ satisfying $|S|\leq m'$, if $U_{m}O_{x}U_{m-1}$$\cdots $ $U_{0}|\psi_{0}\rangle$=$\left[\alpha^{S}_{j,m}\right]$$|F(x)\rangle_{m'}$, then the matrix $\left[\alpha^{S}_{j,m}\right]$ is called the representation matrix of the state $U_{m}O_{x}U_{m-1}$ $\cdots $ $U_{0}|\psi_{0}\rangle$. In order to distinguish the representation matrix of the state $O_{x}U_{m-1}$ $\cdots $ $U_{0}|\psi_{0}\rangle$ from the representation matrix of the state $U_{m}O_{x}U_{m-1}$ $\cdots $ $U_{0}|\psi_{0}\rangle$, the notation $\left[\beta^{S}_{j,m}\right]$ denotes the representation matrix $U^{-1}_{m}\left[\alpha^{S}_{j,m}\right]$ of the state $O_{x}U_{m-1}$ $\cdots $ $U_{0}|\psi_{0}\rangle$.
 \hfill $\Box$
\end{myDef}
Now, Theorem \ref{the2} can be proved as follows.
\begin{proof} $\Rightarrow)$. By using Definition \ref{def6}, the representation matrix of the state $U_{1}O_{x}U_{0}|\psi_{0}\rangle$ can be in the form of $\left[\alpha^{S}_{j,1}\right]=$ $[|\alpha_{\emptyset,1}\rangle,$ $|\alpha_{\{1\},1}\rangle,$ $|\alpha_{\{2\},1}\rangle,$ $\cdots,$ $|\alpha_{\{n\},1}\rangle]$. Meanwhile, since the state in any quantum 1-query algorithm is
$O_{x}U_{0}|\psi\rangle$ $\sum_{j'}\alpha^{\emptyset}_{0,j',0}|0,j'\rangle+$ $\sum_{i,j'}\alpha^{\emptyset}_{i,j',0}(-1)^{x_{i}}|i,j'\rangle,$
the representation matrix of the state $O_{x}U_{0}|\psi_{0}\rangle$ is in the form of $\left[\beta^{S}_{j,1}\right]=$ $diag(B_{0},B_{1},\cdots,B_{n})$ where $B_{i}=$ $(\alpha^{\emptyset}_{i,0,0},$ $\alpha^{\emptyset}_{i,1,0},$ $\cdots,$ $\alpha^{\emptyset}_{i,j',0},\cdots)^{T}$ for all $i\in\{0,1,2,\cdots,n\}$.
Thus, $U^{-1}_{1}$ $\left[\alpha^{S}_{j,1}\right]=$ $diag(B_{0},$ $B_{1},$ $\cdots,$ $B_{n}).$

According to Lemma \ref{lem1}, Definitions \ref{def5} and \ref{def6}, the representation matrix $\left[\alpha^{S}_{j,1}\right]$ of the state $U_{1}O_{x}U_{0}|\psi_{0}\rangle$ can be regarded as an SOS complex representation matrix $[\alpha_{f}]$ of the partial Boolean function $f$ and $\bar{f}$.

Meanwhile, since all columns of any block-diagonal matrix $diag(B_{0},$ $B_{1},$ $\cdots,$ $B_{n})$ are pairwise orthogonal and the unitary operator $U^{-1}_{1}$ preserves the inner product of any two complex vectors, all columns of the matrix $[\alpha_{f}]$ are also pairwise orthogonal. Thus, Eq.~\eqref{reprematrix3} holds.

$\Leftarrow)$. For a degree-1 SOS complex representations matrix $[\alpha_{f}]$ of $f$ and $\bar{f}$ satisfying $[\alpha_{f}]^{\dag}[\alpha_{f}]$ $=$ $diag(u_{0},$ $u_{1},$ $u_{2},$ $\cdots,$ $u_{n})$,
all columns of the matrix $[\alpha_{f}]=$ $[|\alpha_{\emptyset}\rangle,$ $|\alpha_{\{1\}}\rangle,$ $|\alpha_{\{2\}}\rangle,$ $\cdots,$ $|\alpha_{\{n\}}\rangle]$ are pairwise orthogonal.
Note that we can always get a sequence of proper vectors $B_{0},$ $B_{1},$ $\cdots,$ $B_{n}$ satisfying $||B_{0}||$=$|||\alpha_{\emptyset}\rangle||=\sqrt{u_{0}}$ and $||B_{i}||$=$|||\alpha_{\{i\}}\rangle||=\sqrt{u_{i}}$ for all $i\in\{1,2,\cdots,n\}$. As a result, since both all columns of $[\alpha_{f}]$ and all columns of $diag(B_{0},$ $B_{1},$ $\cdots,$ $B_{n})$ are orthogonal bases, there always exists a unitary operator $U^{-1}_{1}$ such that $U^{-1}_{1}$$[\alpha_{f}]$=$diag(B_{0},$ $B_{1},$ $\cdots,$ $B_{n})$.

Based on above analysis, the three states $U_{1}O_{x}U_{0}|\psi_{0}\rangle$, $O_{x}U_{0}|\psi_{0}\rangle$ and $U_{0}|\psi_{0}\rangle$ of an exact quantum 1-query algorithm computing $f$ can be determined by $[\alpha_{f}]$, $diag(B_{0},$ $B_{1},$ $\cdots,$ $B_{n})$ and
\begin{equation}\label{reprematrix1}
\begin{bmatrix}
\begin{smallmatrix}
B_{0}\\
B_{1}\\
 \vdots\\
B_{n}
\end{smallmatrix}
\end{bmatrix},
\end{equation}
respectively. Thus, Theorem \ref{the2} has been proved.
\end{proof}

%In fact, the sufficient and necessary characterization is a $(p+q)\times(1+n)$ complex matrix Eq.~\eqref{representsos} satisfying Eqs.~\eqref{final11} and ~\eqref{reprematrix3}. Here, $S\in\{\emptyset,\{1\},\{2\},\cdots,\{n\}\}$.

\subsection{Proof of Theorem \ref{the1}}\label{proof2}

First, the following lemma is necessary.
\begin{lemma}\label{lem3}
If an $n$-bit partial Boolean function $f: D\rightarrow\{0,1\}$ depends on $k$ ($\leq n$) bits and there exists a degree-1 SOS complex representation of $f$, then there exist at least $k$ non-zero columns $|\alpha_{\{i_{1}\}}\rangle$, $|\alpha_{\{i_{2}\}}\rangle$, $\cdots,$ $|\alpha_{\{i_{k}\}}\rangle$ in the matrix $[\alpha_{f}]$ where $i_{1},$ $ i_{2}, $ $\cdots, $ $i_{k}\in\{1,2,\cdots,n\}$.
\end{lemma}
Then, Theorem \ref{the1} can be proved in the following.
\begin{proof}
On one hand, $G_{f}(1,0)$ is a $(n+1)\times|\{x|f(x)=0\}|$ matrix and $G_{f}(1,1)$ is a $(n+1)\times|\{x|f(x)=1\}|$ matrix.
For a non-constant $n$-bit partial Boolean function $f$, if there exists a degree-1 SOS complex representation, then there exists a sequence of $(n+1)$-dimensional non-zero vectors $|\alpha_{1}\rangle$, $|\alpha_{2}\rangle$, $\cdots,$ $|\alpha_{l}\rangle$ satisfying
\begin{equation}\label{fxsossos}
f(x)=\sum^{p}_{l=0}|(|P(x)\rangle_{1})^{\dag}|\alpha_{l}\rangle|^{2},\forall x\in D.
\end{equation}
Considering Eq.~\eqref{fxsossos} for the truth table 0 of $f$, we know the existence of the sequence (non-zero vectors $|\alpha_{1}\rangle_{1}$, $|\alpha_{2}\rangle_{1}$, $\cdots,$ $|\alpha_{l}\rangle_{1}$) requires
$1\leq(n+1)-$ $rank(G_{f}(1,0))$ $\leq n.$
Similarly, considering a similar Eq.~\eqref{fxsossos} for the truth table 0 of $\bar{f}$, we can get
$1\leq(n+1)$ $-rank(G_{f}(1,1))$ $\leq n$. Thus, Eq.~\eqref{the1equation2} holds.

On the other hand, by the proof of Theorem \ref{the2}, for an $n$-bit partial Boolean function $f$ with exact quantum 1-query complexity, there exists an SOS complex representations matrix $[|\alpha_{\emptyset,1}\rangle,|\alpha_{\{1\},1}\rangle,\cdots,|\alpha_{\{n\},1}\rangle]=[\alpha_{f}]$ of $f$ and $\bar{f}$ such that $U^{-1}_{1}$ $[\alpha_{f}]$ $=diag(B_{0},$ $B_{1},$ $\cdots,$ $B_{n}).$
By Eq.~\eqref{representsos}, we can see that $rank([\alpha_{f}])\leq$ $rank([|\alpha^{1}\rangle,$ $\cdots,|\alpha^{p}\rangle])$$+rank([|\alpha^{p+1}\rangle,$ $\cdots,|\alpha^{p+q}\rangle])$. And, $rank([|\alpha^{1}\rangle,$ $\cdots,|\alpha^{p}\rangle])$ $\leq(1+n)-rank(G_{f}(1,0))$ and $rank([|\alpha^{p+1}\rangle,$ $\cdots,|\alpha^{p+q}\rangle])$ $\leq(1+n)-rank(G_{f}(1,1))$ by considering the truth table 0 of the partial Boolean function $f$ and $\bar{f}$, respectively.
Using the property (preserve Euclidean norm and the rank) of the unitary matrix and Lemma \ref{lem3},
\begin{equation}\label{b1b2b3b4}
\begin{split}
k\leq&|\{i||\alpha_{\{i\},1}\rangle\neq0,i\in\{0,1,\cdots,n\}\}|\\
=&|\{i|B_{i}\neq0,i\in\{0,1,\cdots,n\}\}|\\
=&rank(diag(B_{0},B_{1},\cdots,B_{n}))\\
=&rank([|\alpha_{\emptyset,1}\rangle,|\alpha_{\{1\},1}\rangle,\cdots,|\alpha_{\{n\},1}\rangle])\\
%\leq&rank([|\alpha^{1}\rangle,\cdots,|\alpha^{p}\rangle]){+}rank([|\alpha^{p+1}\rangle,\cdots,|\alpha^{p+q}\rangle])\\
\leq&2(1+n)-rank(G_{f}(1,0))-rank(G_{f}(1,1)).
\end{split}
\end{equation}
%Here, the last inequality holds.
Thus, Eq.~\eqref{the1equation1} can be got and Theorem \ref{the1} has been proved.

\end{proof}

\subsection{Proof of Theorem \ref{the3}}\label{proof3}

In this subsection, let us evaluate the number $N_{1}(n,k)$ of $n$-bit partial Boolean functions which depend on $k$ bits and have exact quantum 1-query complexity.
As a preparation, the following lemma is necessary.
\begin{lemma}\label{lem7}
If $n\geq 2$, for any $j$$(\in\{1,2,\cdots,n+1\})$ different basis vectors $|P(X_{1})\rangle_{1}$, $|P(X_{2})\rangle_{1}$, $\cdots,$ $|P(X_{j})\rangle_{1}$ (these vectors can be chosen from the matrix $G_{n}$), there exist at most $T_{j}\leq2^{j-1}-j$ other different vectors $|P(X_{j+1})\rangle_{1}$, $|P(X_{j+2})\rangle_{1}$, $\cdots,$ $|P(X_{j+T_{j}})\rangle_{1}$ satisfying $rank([|P(X_{1})\rangle_{1},|P(X_{2})\rangle_{1},\cdots,|P(X_{j+T_{j}})\rangle_{1}])=j$.
\end{lemma}
Now, Theorem \ref{the3} can be proved as follows.
\begin{proof}
Let $r_{0}=rank(G_{f}(1,0))$ %$\in\{1,2,\cdots,n\}$
and $r_{1}=rank(G_{f}(1,1))$. % $\in\{1,\cdots,$ $\min\{2n+2-k-r_{0},n\}\}$.
According to Theorem \ref{the1},
$N_{1}(n,k)$ is not bigger than the number of all $n$-bit partial Boolean functions satisfying $r_{0}$ $+r_{1}$ $\leq 2n+2-k$ and $1\leq r_{0},$ $r_{1}\leq n$.
For every fixed $(r_{0},r_{1})$, an $n$-bit partial Boolean function can be determined using the following two steps.

In the first step, we choose $r_{0}$ basis vectors $|P(X_{1})\rangle_{1},$ $|P(X_{2})\rangle_{1},$ $\cdots,$ $|P(X_{r_{0}})\rangle_{1}$ and $r_{1}$ basis vectors $|P(Y_{1})\rangle_{1},$ $|P(Y_{2})\rangle_{1},$ $\cdots,$ $|P(Y_{r_{1}})\rangle_{1}$ from the matrix $G_{n}$, respectively.
Here, $r_{0}$ basis vectors $|P(X_{1})\rangle_{1},$ $|P(X_{2})\rangle_{1},$ $\cdots,$ $|P(X_{r_{0}})\rangle_{1}$ are $r_{0}$ column vectors of the matrix $G_{f}(1,0)$ for an undetermined partial Boolean function $f$, and $r_{1}$ basis vectors $|P(Y_{1})\rangle_{1},$ $|P(Y_{2})\rangle_{1},$ $\cdots,$ $|P(Y_{r_{1}})\rangle_{1}$ are $r_{1}$ column vectors of the matrix $G_{f}(1,1)$ for the undetermined partial Boolean function $f$.
For every fixed $(r_{0},r_{1})$, the number of different selections (i.e., $\{|P(X_{1})\rangle_{1},$ $|P(X_{2})\rangle_{1},$ $\cdots,$ $|P(X_{r_{0}})\rangle_{1}\}$ and $\{|P(Y_{1})\rangle_{1},$ $|P(Y_{2})\rangle_{1},$ $\cdots,$ $|P(Y_{r_{1}})\rangle_{1}\}$) is not bigger than
\begin{equation}\label{eq21roriupper}
\begin{split}
\binom{2^{n}}{r_{0}}\binom{2^{n}-r_{0}}{r_{1}}
\leq2^{n(r_{0}+r_{1})}
\end{split}
\end{equation}
for all $n\geq3$ and $k\geq 2$.

In the second step, we add some other vectors $|P(X_{r_{0}+1})\rangle_{1},$ $\cdots$ and $|P(Y_{r_{1}+1})\rangle_{1},$ $\cdots$ to the set $\{|P(X_{1})\rangle_{1},$ $|P(X_{2})\rangle_{1},$ $\cdots,$ $|P(X_{r_{0}})\rangle_{1}\}$ and the set $\{|P(Y_{1})\rangle_{1},$ $|P(Y_{2})\rangle_{1},$ $\cdots,$ $|P(Y_{r_{1}})\rangle_{1}\}$, respectively.
Here, every newly added vector in the set $\{|P(X_{r_{0}+1})\rangle_{1},$ $\cdots\}$ should be represented linearly by the determined $r_{0}$ basis vectors $|P(X_{1})\rangle_{1},$ $|P(X_{2})\rangle_{1},$ $\cdots,$ $|P(X_{r_{0}})\rangle_{1}$
and every newly added vector in the set $\{|P(Y_{r_{1}+1})\rangle_{1},$ $\cdots\}$ should be represented linearly by the determined $r_{1}$ basis vectors $|P(Y_{1})\rangle_{1},$ $|P(Y_{2})\rangle_{1},$ $\cdots,$ $|P(Y_{r_{1}})\rangle_{1}$.
After that, an $n$-bit partial Boolean function $f$ is determined as follows.
$f(x)=0$ for $x$ in the set $\{X_{1},X_{2},$ $\cdots,X_{r_{0}},\cdots\}$, $f(x)=1$ for $x$ in the set $\{Y_{1},Y_{2},$ $\cdots,Y_{r_{1}},\cdots\}$, and it is undefined for the rest cases.
%Without loss of generality, assume that $k\geq 2$.
Using Eq.~\eqref{eq21roriupper} and Lemma \ref{lem7}, for every fixed $(r_{0},r_{1})$,
there are at most
\begin{equation}\label{upperbound1}
\begin{split}
&2^{n(r_{0}+r_{1})}2^{(2^{r_{0}-1}-r_{0})}2^{(2^{r_{1}-1}-r_{1})}\\
<&2^{2n^{2}}2^{(2^{n-1}+2^{n+1-k})}
=2^{2^{n-1}(1+2^{2-k})+2n^{2}}
\end{split}
\end{equation}
partial Boolean functions with $rank(G_{f}(1,0))$ $=r_{0}$ and $rank(G_{f}(1,1))$ $=r_{1}$.
Note that the number of different $(r_{0},r_{1})$ is not bigger than $n^{2}$.
Thus, Theorem \ref{the3} has been proved.
\end{proof}

\section{CONCLUSIONS}\label{Concl}

In this paper, we have investigated the power and advantage of the exact quantum 1-query model for partial Boolean functions. Specifically, we have contributed two sufficient and necessary  conditions for characterizing $n$-bit partial Boolean functions with exact quantum 1-query complexity, and one necessary condition for characterizing $n$-bit partial Boolean functions that depend on $k$ $(k\leq n)$ bits and have exact quantum 1-query complexity.
Using these characterizations, we have clarified all  $n$-bit partial Boolean functions that depend on $n$ bits with exact quantum 1-query complexity (in fact, $n\leq 2$ in this case, i.e. Theorem 2). Also, we have proved that the number of all $n$-bit partial Boolean functions that depend on $k$ ($k\leq n$) bits with exact quantum 1-query complexity is quite small.
As a result, the following two problems are worthy of further consideration.
\begin{itemize}
\item [(1)]~{\bf Find all (or some) non-trivial $n$-bit partial Boolean functions with exact quantum 1-query complexity.} This is an interesting problem for the following two aspects.
    On one hand, the upper bound (given by this paper) of the actual number of partial Boolean functions in this class is quite big. On the other hand, known non-trivial $n$-bit partial Boolean functions in this class are still fairly rare.
\item [(2)]~{\bf How many $n$-bit partial Boolean functions can be computed exactly (or with bounded-error) by exact quantum $k$-query algorithms for all $k\in\{2,3,\cdots,n\}$?} The solution of this problem is a quantitative evaluation of the advantage of the quantum $k$-query model. In contrast, the result of Ambainis et al. \cite{Ambainis2015qic} is a qualitative evaluation of the advantage of the quantum query model.
\end{itemize}

\section*{Acknowledgements}
This work is supported in part by the National Natural Science Foundation of China (Nos. 61572532, 61876195), the Natural Science Foundation of Guangdong Province of China
(No. 2017B030311011).

\appendix

\section{The background of Definition \ref{def441}}
First, the following definition is used widely for total Boolean functions (i.e., $D=\{0,1\}^{n}$) \cite{Simon1983A,Ambainis2015qic,Chen2020character}.
\begin{myDef}\label{defdepend}
\cite{Simon1983A,Chen2020character}. Given an $n$-bit partial Boolean function $f: D\rightarrow\{0,1\}$, we say that $f$ depends on the $k$-th ($k\in\{1,2,\cdots,n\}$) bit if there exists a pair of inputs $X_{k},X^{\{k\}}_{k}\in D$ such that $1\oplus f(X_{k})=f(X^{\{k\}}_{k})\in\{0,1\}$.
Here, $X^{\{k\}}_{k}$ is the same as $X_{k}$ except for the $k$-th bit being flipped.
%Thus, if an $n$-bit partial Boolean function $f: D\rightarrow\{0,1\}$ does not depend on the $k$-th ($k\in\{1,2,\cdots,n\}$) bit, then $1\oplus f(X_{k})=f(X^{\{k\}}_{k})\in\{0,1\}$ holds for any pair of inputs $X_{k},X^{\{k\}}_{k}\in D$.
\hfill $\Box$
\end{myDef}
Clearly, if a total Boolean function $f: D=\{0,1\}^{n}\rightarrow\{0,1\}$ depends on $k$ bits, then we can always find out a sequence $i_{1}$, $i_{2}$, $\cdots$, $i_{k}$ such that $f$ depends on the $i_{1}$-bit, the $i_{2}$-bit, $\cdots,$ and the $i_{k}$-bit. Meanwhile, the unique multilinear polynomial representation of $f$ is on the $k$ variables (i.e., $x_{i_{1}}$, $x_{i_{2}}$, $\cdots,$ $x_{i_{k}}$).
However, for $D\neq\{0,1\}^{n}$, things become different.
On one hand, the even $n$-bit partial Boolean function $g$ computed by Deutsch-Jozsa algorithm \cite{Deutsch1992rapidsolution} does not depend on any bit (using Definition \ref{defdepend}).
On the other hand, any multilinear polynomial representation of $g$ is on at least $\frac{n}{2}+1$ variables.
This typical partial Boolean function $g$ motivates us to use Definition \ref{def441} in this paper.
Specifically, in the case of total Boolean functions, Definition \ref{def441} is  consistent with Definition \ref{defdepend}.

\section{Proof of Lemma \ref{lem8}}

\begin{proof}
First, for any $k$-bit $(k\geq 1)$ partial Boolean function $f: D\rightarrow \{0,1\}$ ($D\subseteq\{0,1\}^{k}$) depending on $k$ bits, there exists at least one input $y=y_{1}y_{2}\cdots y_{k}\in\{0,1\}^{k}$ such that $f(y)\in\{0,1\}$. Then, any $(k+1)$-bit partial Boolean function $g$ defined by
\begin{equation}\label{constructanewg}
\left \{
 \begin{aligned}
&g(x0)=f(x), &\forall x\in D,\\
&g(x0)=*, &\forall x\in\{0,1\}^{k}/D,\\
&g(y1)=1\oplus f(y), & x=y,\\
&g(x1)\in\{0,1,*\}, &\forall x\in\{0,1\}^{k}/\{y\}
 \end{aligned}
\right.
\end{equation}
is a $(k+1)$-bit partial Boolean function depending on $(k+1)$ bits. Thus, for any $k\geq 1$,
\begin{equation}\label{}
\frac{N(k+1)}{N(k)}\geq3^{2^{k}-1}.
\end{equation}
Since $N(1)=2$ (i.e., $(f(0),$ $f(1))=$ $(0,1)$ and $(f(0),$ $f(1))=$ $(1,0)$), we have
\begin{equation}\label{}
N(n)\geq2\prod^{n-1}_{k=1}3^{2^{k}-1}=2\times3^{2^{n}-n-1}.
\end{equation}
The lemma has been proved.
\end{proof}

\section{Proof of Lemma \ref{lem1}}

\begin{proof} By Lemma 7 in  \cite{Buhrman2002Complexity}, for every quantum 1-query algorithm computing an $n$-bit partial Boolean function $f: D\rightarrow\{0,1\}$, the final state $U_{1}O_{x}U_{0}|\psi_{0}\rangle$ is in the form of $\sum_{j}\alpha_{j}(x)|j\rangle$ where every $\alpha_{j}(x)$ is a complex-valued $n$-variate multilinear polynomial and $\deg(\alpha_{j}(x))\leq 1$ for any $j$.
According to  Theorem 17 in  \cite{Buhrman2002Complexity}, the acceptance probability of the set of basis states corresponding to a 1-output is
\begin{equation}\label{assignf}
\sum_{j:\textrm{Output}~1}|\alpha_{j}(x)|^{2}=f(x)~\textrm{for~all}~x\in D.
\end{equation}
Since $f$ is not a constant, Eq.~\eqref{assignf} is a degree-1 SOS complex representation of $f$.

Since the final state $\sum_{j}\alpha_{j}(x)|j\rangle$ is a unit vector and the algorithm outputs $f(x)$ exactly (i.e., the algorithm outputs 1 with probability $f(x)$ and outputs 0 with probability $1-f(x)$ for all $x\in D$), the acceptance probability of the set of basis states corresponding to a 0-output is
\begin{equation}\label{assign1-f}
\sum_{j:\textrm{Output}~~0}|\alpha_{j}(x)|^{2}=\bar{f}(x)~\textrm{for~all}~x\in D.
\end{equation}
Eq.~\eqref{assign1-f} is an SOS complex representation of $\bar{f}$. The lemma has been proved.
\end{proof}

\section{Proof of Lemma \ref{lem3}}
\begin{proof} If there exist $n-k+1$ zero vectors $|\alpha_{\{i_{k}\}}\rangle,$ $|\alpha_{\{i_{k+1}\}}\rangle,$ $\cdots,$ $|\alpha_{\{i_{n}\}}\rangle$ where $i_{k}$, $i_{k+1}$, $\cdots,$ $i_{n}$ $\in\{1,2,\cdots,n\}$, then all entries $\alpha^{l}_{\{i_{r}\}}$ in $|\alpha_{\{i_{r}\}}\rangle$ are zeros for all $l\in\{1,2,\cdots,p+q\}$ and $r\in\{k,k+1,\cdots,n\}$. Therefore,
\begin{equation}\label{}
 \begin{split}
f(x)=&|f_{1}(x)|^{2}+\cdots+|f_{p}(x)|^{2}\\
=&\sum^{p}_{l=1}\left|\alpha^{l}_{\emptyset}+\sum_{i\in\{1,2,\cdots,n\}}\alpha^{l}_{\{i\}}(-1)^{x_{i}}\right|^{2}\\
=&\sum^{p}_{l=1}\left|\alpha^{l}_{\emptyset}+\sum_{r\in\{1,2,\cdots,k-1\}}\alpha^{l}_{\{i_{r}\}}(-1)^{x_{i_{r}}}\right|^{2}.
 \end{split}
\end{equation}
Obviously, $f$ depends on at most $k-1$ bits. This is a contradiction. Thus, the lemma has been proved.
\end{proof}

%\section{Proof of Lemma \ref{lem6}}
%\begin{proof} If there exist $k_{1}, k_{2}$ such that $|P(x)\rangle_{1}=k_{1}|P(y)\rangle_{1}+k_{2}|P(z)\rangle_{1}$, then $k_{1}+k_{2}=1$ and $k_{1}y_{i}+(1-k_{1})z_{i}=x_{i}$ for all $i\in\{1,2,\cdots,n\}$. Since $y\neq z$, there must exist an $i$ such that $(y_{i},z_{i})=(0,1)$ or $(y_{i},z_{i})=(1,0)$. Any of them implies $k_{1}\in\{0,1\}$. If $k_{1}=0$, then we have $z=x$. Otherwise, $y=x$. This is a contradiction. Thus, the lemma has been proved.
%\end{proof}

\section{Proof of Lemma \ref{lem7}}
\begin{proof} For every $k$ $\geq $ $j+1$ and input $X_{r}=X_{r,1}X_{r,2}$ $\cdots $ $X_{r,n}\in\{0,1\}^{n}$ where $r\in\{1,2,\cdots,j,k\}$, let $|P(X_{k})\rangle_{1}$ $=$ $s_{1}|P(X_{1})\rangle_{1}$ + $s_{2}|P(X_{2})\rangle_{1}$ + $\cdots$
+ $s_{j}|P(X_{j})\rangle_{1}$ which is an equation set on $j$ variables $s_{1}$, $s_{2}$, $\cdots,$ $s_{j}$. Note that this equation set
\begin{equation}\label{equati12}
\left \{
 \begin{aligned}
&\sum^{j}_{r=1}s_{r}=1, &\\
&\sum^{j}_{r=1}s_{r}X_{r,i}=X_{k,i}, &i\in\{1,2,\cdots,n\}.
 \end{aligned}
\right.
\end{equation}
consists of $n+1$ equations. Since vectors $|P(X_{1})\rangle_{1}$, $|P(X_{2})\rangle_{1}$, $\cdots,$ $|P(X_{j})\rangle_{1}$ is a base, the rank of the matrix ($|P(X_{1})\rangle_{1}$, $|P(X_{2})\rangle_{1}$, $\cdots,$ $|P(X_{j})\rangle_{1}$) is $j$. In other word, for all rows of the matrix ($|P(X_{1})\rangle_{1}$, $|P(X_{2})\rangle_{1}$, $\cdots,$ $|P(X_{j})\rangle_{1}$), we can find out a base which consists of $j$ row vectors. After that, the augmented matrix of the equation set Eq.~\eqref{equati12}
\begin{equation}\label{augment000}
\begin{bmatrix}
1 & 1 & \cdots& 1 & 1\\
X_{1,1}  & X_{2,1} &\cdots & X_{j,1} & X_{k,1}\\
X_{1,2}  & X_{2,2} &\cdots & X_{j,2} & X_{k,2}\\
\vdots & \vdots & & \vdots & \vdots \\
X_{1,n}  & X_{2,n} &\cdots & X_{j,n} & X_{k,n}
\end{bmatrix}
\end{equation}
can be transformed into
\begin{equation}\label{augment}
\begin{bmatrix}
1 & 1 & \cdots& 1 & 1\\
X_{1,i_{1}}  & X_{2,i_{1}} &\cdots & X_{j,i_{1}} & X_{k,i_{1}}\\
X_{1,i_{2}}  & X_{2,i_{2}} &\cdots & X_{j,i_{2}} & X_{k,i_{2}}\\
\vdots & \vdots & & \vdots & \vdots \\
X_{1,i_{j-1}}  & X_{2,i_{j-1}} &\cdots & X_{j,i_{j-1}} & X_{k,i_{j-1}}\\
0  & 0 &\cdots & 0 & X'_{k,i_{j}}\\
0  & 0 &\cdots & 0 & X'_{k,i_{j+1}}\\
\vdots & \vdots & & \vdots & \vdots \\
0  & 0 &\cdots & 0 & X'_{k,i_{n}}
\end{bmatrix}
\end{equation}
where $\{i_{1},i_{2},\cdots,i_{n}\}$ $=\{1,2,\cdots,n\}$.
According to the solution theory of linear system
of equations, if any of $X'_{k,i_{j}}$, $X'_{k,i_{j+1}}$, $\cdots,$ and $X'_{k,i_{n}}$ is non-zero, then there does not exist a solution of the equation set Eq.~\eqref{equati12} and the vector $|P(X_{k})\rangle_{1}$ is not what we want. Otherwise, we can always get a solution
\begin{equation}\label{}
\begin{bmatrix}
s_{1}\\
s_{2}\\
s_{3}\\
\vdots\\
s_{j}
\end{bmatrix}=\begin{bmatrix}
1 & 1 & \cdots& 1\\
X_{1,i_{1}}  & X_{2,i_{1}} &\cdots & X_{j,i_{1}}\\
X_{1,i_{2}}  & X_{2,i_{2}} &\cdots & X_{j,i_{2}}\\
\vdots & \vdots & & \vdots\\
X_{1,i_{j-1}}  & X_{2,i_{j-1}} &\cdots & X_{j,i_{j-1}}
\end{bmatrix}^{-1}
\begin{bmatrix}
1\\
X_{k,i_{1}}\\
X_{k,i_{2}}\\
\vdots \\
X_{k,i_{j-1}}
\end{bmatrix}.
\end{equation}

As a result, for every $X_{k,i_{1}}X_{k,i_{2}}\cdots X_{k,i_{j-1}}\in\{0,1\}^{j-1}$, we either can get a unique string $X_{k}=$ $X_{k,1}X_{k,2}\cdots X_{k,n}\in\{0,1\}^{n}$ satisfying $X'_{k,i_{j}}=$ $X'_{k,i_{j+1}}$= $\cdots$ =$X'_{k,i_{n}}=0$ in Eq.~\eqref{augment} or can not get a string $X_{r,1}X_{r,2}\cdots X_{r,n}\in\{0,1\}^{n}$ satisfying $X'_{k,i_{j}}=$ $X'_{k,i_{j+1}}$= $\cdots$ =$X'_{k,i_{n}}=0$ in Eq.~\eqref{augment}. The lemma has been proved.
\end{proof}

\nocite{*}
%\bibliography{apssamp}% Produces the bibliography via BibTeX.

\end{document}